\documentclass[%
 reprint,superscriptaddress,
 amsmath,amssymb,
]{revtex4-1}
\usepackage{natbib}
\usepackage{graphicx}
\usepackage{dcolumn}
\usepackage{bm}
\usepackage{color}
\usepackage{enumerate}
\usepackage{enumitem}
\usepackage[normalem]{ulem}
\usepackage{lipsum}
\usepackage{subfigure}

\usepackage{xcolor, soul}
\sethlcolor{yellow}

\makeatletter
\AtBeginDocument{\let\LS@rot\@undefined}
\makeatother

\begin{document}

\title{Spin crossover transition driven by pressure: Barocaloric applications}

\author{Mario Reis} \email{marioreis@id.uff.br}
\affiliation{Institute of Physics, Fluminense Federal University, Av. Gal. Milton Tavares de Souza s/n, 24210-346, Niteroi-RJ, Brazil}
\affiliation{Dpto. F\'{i}sica de la Materia Condensada, Universidad de Sevilla, Apdo 1065, 41080 Sevilla, Spain}

\author{Yongqiang Cheng}
\email{chengy@ornl.gov}
\affiliation{Neutron Scattering Division, Oak Ridge National Laboratory, Oak Ridge, TN 37831-6475 United States}

\author{Antonio M. dos Santos}
\email{dossantosam@ornl.gov}
\affiliation{Neutron Scattering Division, Oak Ridge National Laboratory, Oak Ridge, TN 37831-6475 United States}


\begin{abstract}

This article describes a mean-field theoretical model for Spin-Crossover (SCO) materials and explores its implications. It is based on a simple Hamiltonian that yields the high spin molar fraction as a function of temperature and pressure, as well as a temperature-pressure phase diagram for the SCO transition. In order to test the model, we apply it to the giant Barocaloric Effect (BCE) of the SCO material [FeL$_2$][BF$_4$]$_2$ and comprehensively analyse its behavior. We found that optical phonons are responsible for 92\% of the total barocaloric entropy change. DFT calculations show that these optical phonons are mainly assigned to the low frequencies modes of vibration ($<400$ cm$^{-1}$), being associated to the Fe coordination.

\end{abstract}

\keywords{}
\maketitle

\section{Introduction}

Spin-Crossover (SCO) materials exhibit a transition between two spin states, that may be driven by pressure, temperature or radiation. Depending on the system, the crossover temperature may be extremely sensitive to the external excitation. This sensitivity makes them particularly useful for applications in devices, including switches, actuators, data storage and many others\cite{coronado2020molecular}. It was recently proposed by  Sandeman\cite{sandeman2016research}, that the combination of a large volume change across the SCO transition coupled with a pronounced sensitivity to pressure could make SCO compounds promising candidates to exhibit strong Barocaloric Effect (BCE). This prediction was confirmed in the molecular compound [FeL$_2$][BF$_4$]$_2$, where L=2,6-di(pyrazol-1-yl)pyridine (BPP)\cite{vallone2019giant}. More recently, this BCE has also been theoretically explored by Ribeiro\cite{ribeiro2019influence}, von Ranke\cite{von2021refrigeration,von2020large} and co-workers. 

The present article proposes an extension of the SCO model developed by Babilotte and Boukheddaden\cite{babilotte2020theoretical}, that allows the creation of a detailed temperature-pressure phase diagram mapping the SCO transition.  From those results, the barocaloric entropy change could be obtained and the mean-field thermodynamic model successfully deployed. Density functional theory (DFT) provided important molecular dynamic parameters - inaccessible from previous diffraction data. This parametrization of the proposed model, combining theory and experimental details about the material, led to a satisfactory match between theory and experiment with minimal free parameters. This model is tested against the observation of giant BCE on a SCO compound [FeL$_2$][BF$_4$]$_2$ that was described in an earlier work \cite{vallone2019giant}.

\section{Mean-field model for SCO transition: pressure effect and phase diagram}

\subsection{The model}\label{tmodel}
The Hamiltonian used to model the SCO behavior follows what was proposed earlier by Babilotte \cite{babilotte2020theoretical}:
\begin{equation}
    H=\frac{K}{2}\sum_{ij}(x_{ij}-a_{ij})^2+\Delta_{eff}\sum_i\sigma_i+P\sum_{ij}x_{ij}.
\end{equation}
The first term accounts for the lattice elasticity, and the second term for the energy gap between the $e_g$ and $t_{2g}$ orbitals, where $\sigma=\pm 1$ represents a virtual spin, being $+1$ a HS state and $-1$ a LS state. Finally, the third term incorporates the effect of applied pressure. Considering an homogeneous distance about the metal centers, one can simplify $x_{ij}=x$. Also following the rational presented in  reference \cite{babilotte2020theoretical}, the operator:
\begin{equation}\label{aij}
    a_{ij}=a_{HL}+\frac{\delta a}{4}(\sigma_i+\sigma_j)
\end{equation}
is defined. Here $\delta a=a_{HH}-a_{LL}$ represents the lattice parameter difference between HS-LS states; and $a_{HH}$, $a_{LL}$, $a_{LH}=a_{HL}$ are the corresponding metal-metal atomic distances. 

Considering equation \ref{aij}, the proposed Hamiltonian describing the SCO mechanism can be rewritten as:
\begin{equation}\label{H2}
    H=\frac{k}{2}(x-a_{HL})^2+px+h\sum_i\sigma_i+J\sum_{ij}\sigma_i\sigma_j,
\end{equation}
where:
\begin{equation}
    k=KN\frac{q}{2},\;\;\;\;\;\;\;p=PN\frac{q}{2},\;\;\;\;\;\;\;J=K\left(\frac{\delta a}{4}\right)^2
\end{equation}
and
\begin{equation}
h=\Delta_{eff}-K\frac{q}{2}\left(\frac{\delta a}{2}\right)(x-a_{HL)}.
\end{equation}

Above, we have considered:
\begin{equation}
    \sum_{ij}=\sum_{i=1}^N\sum_{j\in nn(i)}=\frac{q}{2}N.
\end{equation}
In other words, the sum runs over all $i$ metal sites, from 1 up to $N$; and for each site $i$, there is another sum, counting the next-nearest neighbors, i.e., $nn(i)$. We are considering $q$ next-nearest neighbors and dividing the result by two to avoid double counting. The Hamiltonian described in equation \ref{H2} also contains an extra term, given by $D\sum_i\sigma_i^2$, where $D=Jq/2$.  This contribution represents a local anisotropy and therefore can be disregarded, considering the last term of equation \ref{H2} is a Ising-like term.

Equation \ref{H2} can also be written by applying a mean field approach by considering $\sigma_i=\langle\sigma_i\rangle+\delta\sigma_i$, where $\langle\sigma_i\rangle$ is the mean value for $\sigma_i$ and $\delta\sigma_i$ the corresponding fluctuation. From this assumption, neglecting $\delta\sigma_i\delta\sigma_j$ terms, we obtain:
\begin{equation}
    \sigma_i\sigma_j=m(\sigma_i+\sigma_j)-m^2,
\end{equation}
where $m=\langle\sigma_l\rangle$ represents the virtual magnetization for this Ising model. The mean-field Hamiltonian for the SCO mechanism then reads as:
\begin{equation}\label{HMF}
    H_{MF}=E_p^\prime-h_{eff}\sum_i\sigma_i,
\end{equation}
where
\begin{equation}
    h_{eff}=-h-Jqm
\end{equation}
and
\begin{equation}
    E_p^\prime=\frac{k}{2}(x-a_{HL})^2+px-jm^2.
\end{equation}
Above, $j=JNq/2$. From the virtual magnetization $m$, it is possible to obtain the HS molar fraction $n_{HS}$ and the LS molar fraction $n_{LS}=1-n_{HS}$. In this description, for  $m=+1$ ($m=-1$) all molecules are on the HS (LS) state. It is then possible to write:
\begin{equation}\label{nhseq}
    n_{HS}=\frac{(m+1)}{2}.
\end{equation}
As expected, this molar fraction can vary between zero and one.


\subsection{Equilibrium thermodynamics}

From equation \ref{HMF}, we can obtain the partition function for this model:
\begin{align}\nonumber
    Z_{MF}&=\sum_{\sigma_i=\pm 1}e^{-\beta H_{MF}}\\
    &=e^{-\beta E_p^\prime}\left[2\cosh\left(\beta h_{eff}\right)\right]^N,
\end{align}
where $\beta=1/k_BT$. The virtual magnetization $m=\langle\sigma_i\rangle$ can be written as:
\begin{align}\label{mself}\nonumber
    m&=\frac{1}{NZ_{MF}}\sum_{\sigma_i=\pm 1}\sigma_ie^{-\beta H_{MF}}\\
    \nonumber&=\frac{1}{N\beta}\frac{\partial}{\partial h_{eff}}\ln(Z_{MF})\\
    &=\tanh(\beta h_{eff}).
\end{align}
From these results, the free energy can be written as:
\begin{align}\nonumber
    F_{MF}&=-k_BT\ln(Z_{MF})\\
    &=E_p^\prime-k_BTN\ln\left[2\cosh\left(\beta h_{eff}\right)\right].
\end{align}

The equilibrium position, based on the condition
\begin{equation}
    \left.\frac{\partial F_{MF}}{\partial x}\right|_{x=\tilde{x}}=0
\end{equation}
is
\begin{equation}
    (\tilde{x}-a_{HL})=m\frac{\delta a}{2}-\frac{p}{k}.
\end{equation}
From this equilibrium position, the equilibrium free energy $\tilde{F}_{MF}$ can be obtained:
\begin{equation}\label{Feqqq}
    \tilde{F}_{MF}=\tilde{E}_p^\prime-k_BTN\ln\left[2\cosh\left(\beta \tilde{h}_{eff}\right)\right],
\end{equation}
where
\begin{equation}
    \tilde{E}_p^\prime=-\frac{p^2}{2k}+pa_{HL}+jm^2;
\end{equation}
and the equilibrium effective field: 
\begin{equation}\label{heff3}
    \tilde{h}_{eff}=h_0+k_BT_0m.
\end{equation}
Above,
\begin{equation}\label{t0frec3}
    h_0=-\Delta_{eff}-P^\prime,\;\;\;\;\;P^\prime=\delta a\frac{q}{4}P\;\;\;\;\text{and}\;\;\;\;\;k_BT_0=Jq.
\end{equation}
Considering $\delta a$ represents the lattice parameter difference between HS-LS states and $q$ the number of next-nearest neighbors, we can assume $q\delta a=\delta V$, where $\delta V$ is the volume change across the SCO transition.

SCO transition can be induced by external excitation, such as temperature, mechanical pressure and radiation \cite{coronado2020molecular}. Thus, the effective energy splitting can be considered as the ligand-field energy $\Delta_0$ minus an amount of entropy $\Delta S$, due to the spin crossover transition. This amount of entropy is related with an amount of heat, written as:
\begin{align}
    \nonumber\Delta Q&=T\Delta S\\
    \nonumber&=T(S_{HS}-S_{LS})\\
    &=k_BT(\ln g_{HS}-\ln g_{LS}),
\end{align}
where $g_i$ are degeneracies . Thus, the effective energy splitting can be written as\cite{babilotte2020theoretical}:
\begin{equation}\label{Deff}
    \Delta_{eff}=\Delta_0-k_BT\ln(g),
\end{equation}
where $g_{HS}$ and $g_{LS}$ are the degeneracies of the corresponding state, and $g=g_{HS}/g_{LS}$ stands for the degeneracy ratio. For the present study, and many others\cite{babilotte2020theoretical,von2017microscopic}, $g$ is a number; however, Ribeiro \cite{ribeiro2019influence} and von Ranke\cite{von2022theoretical} have been considering a more complex dependence for $g$.

At low temperatures, $\Delta_0\gg k_BT\ln(g)$ and the effective crystal field splitting is bigger than the electron pairing energy. Thus, the system lies in the LS state. Increasing the temperature, the effective crystal field splitting decreases in value and eventually becomes comparable with the electron pairing energy. As a consequence, the system changes to a HS state upon further increase of temperature. 

Replacing equation \ref{Deff} into \ref{heff3}, the effective field can be rewritten as:
\begin{equation}\label{h0eff}
    \tilde{h}^0_{eff}=\frac{\tilde{h}_{eff}}{k_BT_0}=-\gamma+t\ln(g)+m,
\end{equation}
where
\begin{equation}\label{gamma}
    \gamma=\frac{\Delta_0+P^\prime}{k_BT_0}\;\;\;\;\;\text{and}\;\;\;\;\;t=\frac{T}{T_0}.
\end{equation}

\begin{figure}
	\centering
	\includegraphics[width=\columnwidth,keepaspectratio]{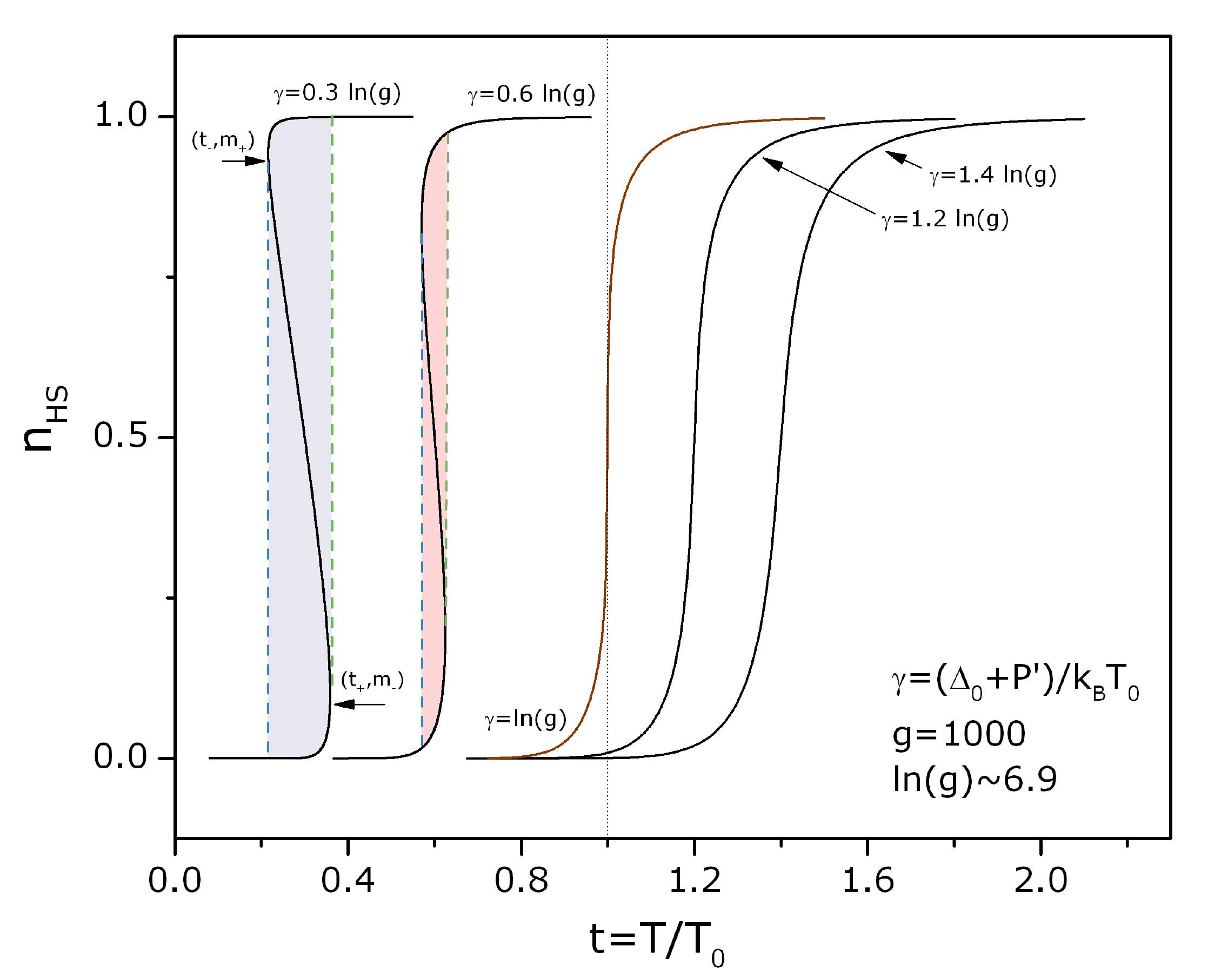}
	\caption{High spin molar fraction $n_{HS}$ as a function of reduced temperature $t$. These results were obtained from the Ising-like model described in section \ref{tmodel}. The condition for hysteresis is $\gamma<\ln(g)$.\label{nhs}}
\end{figure}

Based on equation \ref{h0eff}, the inverse function for the virtual magnetization is:
\begin{equation}\label{tmm1}
    t=\frac{2(m-\gamma)}{\ln\left[\frac{(1+m)}{g^2(1-m)}\right]}.
\end{equation}
For $m=0$, the fraction of molecules in the HS state is $n_{HS}=1/2$ and therefore:
\begin{equation}\label{tdem}
    t_{eq}=t(m=0)=\frac{\gamma}{\ln(g)}.
\end{equation}
Figure \ref{nhs} presents the fraction of HS molecules $n_{HS}$ (see equation \ref{nhseq}), as a function of the reduced temperature $t=T/T_0$, for some values of $\gamma$ (representing different values of pressure - see equation \ref{gamma}). Note that, for $\gamma>\ln(g)$, there is no hysteresis, with $t_{eq}$ being the temperature in which there is an inflection point for $n_{HS}(T)$; while for $\gamma<\ln(g)$, there is hysteresis. 

\subsection{Phase diagram}

\begin{figure}
	\centering
	\includegraphics[width=\columnwidth,keepaspectratio]{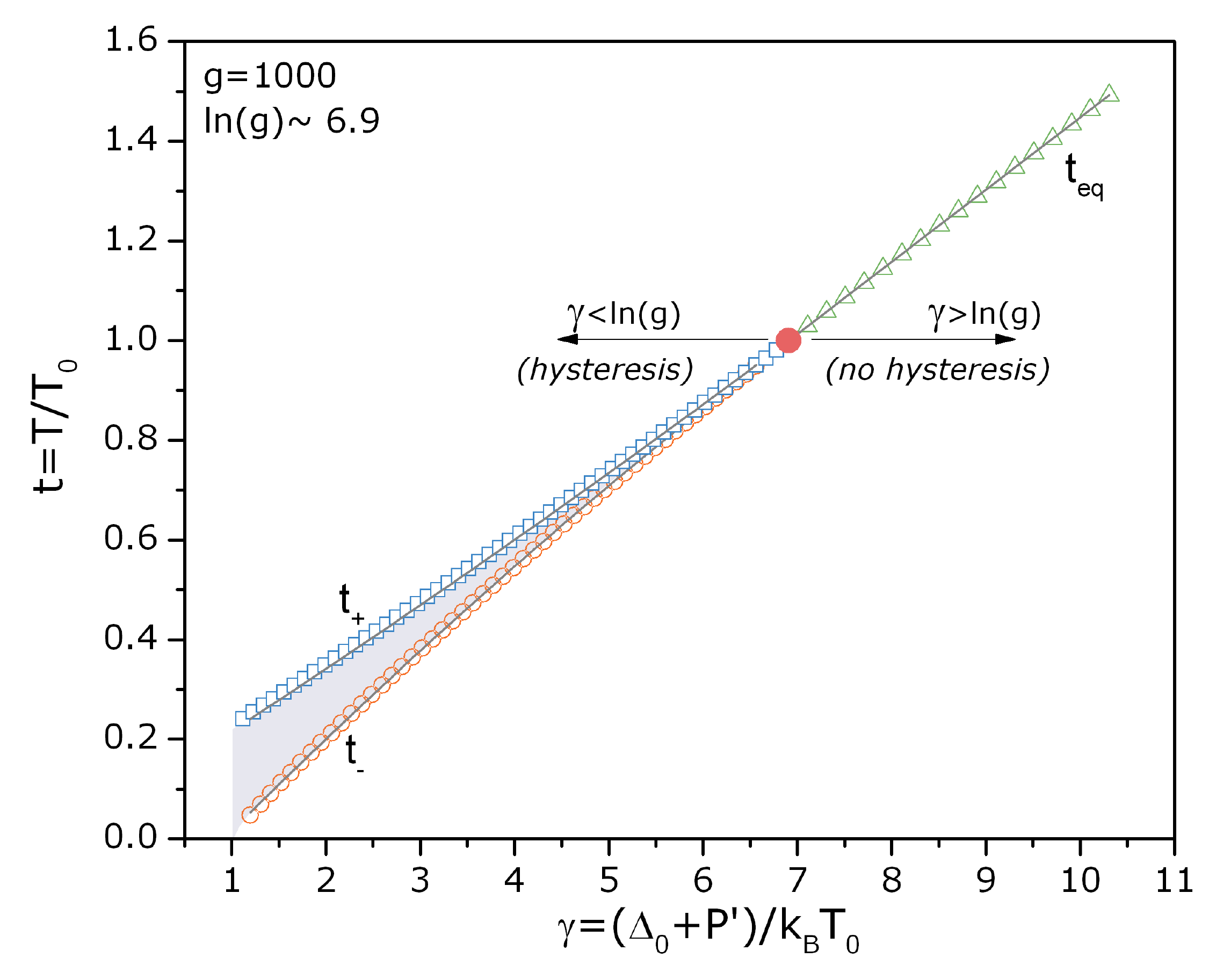}
	\caption{Temperature-Pressure phase diagram. $1<\gamma<\ln(g)$ is the region with hysteresis and $\gamma>\ln(g)$ the region without hysteresis.\label{PD}}
\end{figure}

For the condition $\gamma<\ln(g)$, there are two specific temperatures to explore: $t_+(m_-)$ and $t_-(m_+)$, where $t_+>t_-$ and $m_+>m_-$. In order to obtain further information about these specific temperatures, we must find the minimum of the equation \ref{tmm1}, i.e., $\frac{dt}{dm}|_{m_\pm}=0$:
\begin{equation}\label{fwf}
    2m_\pm\left[1-m_\pm\ln(g)\right]+(m_\pm^2-1)\ln\left(\frac{1+m_\pm}{1-m_\pm}\right)=2\left[\gamma-\ln(g)\right].
\end{equation}
The equation (\ref{fwf}) above was solved numerically for $m_\pm$ for a set of $\gamma$ and for $g=1000$.\cite{von2017microscopic} The results were plugged into equation \ref{tmm1} to obtain the specific values $t_\mp(m_\pm)$, as shown in the phase diagram in Figure \ref{PD}. 

It is possible to determine a minimum value for $\gamma$. The minimum value for $t_-$ is $t_-^{min}=0$, as temperature is always positive. From equation \ref{tmm1}, it is possible to obtain $m_+^{max}=1$, following from $t_-^{min}=0$. The condition $m_+^{max}=1$ into equation \ref{fwf} gives the unity as the minimum value for $\gamma$. Thus, hysteresis will occur once the following condition is true: $1<\gamma<\ln(g)$. 

As can be seen in Figure \ref{PD}, hysteresis is predicted for $1<\gamma<\ln(g)$; and no hysteresis is observed for larger $\gamma$ values. These regions are directly related to the value of applied pressure, as $\gamma$ depends on pressure  (see equation \ref{gamma}). However, one question arises: what is the influence of $g$ on the onset of hysteresis? To answer this question, a similar procedure was performed  (not shown), for $10^2\lesssim g \lesssim 10^4$; and a polynomial fit were made on $t_\pm(\gamma)$ branches using the equation:
\begin{align}\label{tmamee}\nonumber
    t_\pm(\gamma)&=C_0^\pm+
    C_1^\pm\gamma+C_2^\pm\gamma^2\\
    &=\sum_{n=0}^2C_n^\pm\gamma^n.
\end{align}
One fit can be seen on the hysteresis region of the PD in Figure \ref{PD}. The parameters $C_n^\pm$ are in Figure \ref{para}, as a function of $\ln(g)$, and the fit to the numerical results (dots), was made empirically, following the function:
\begin{figure}
	\centering
	\includegraphics[width=\columnwidth,keepaspectratio]{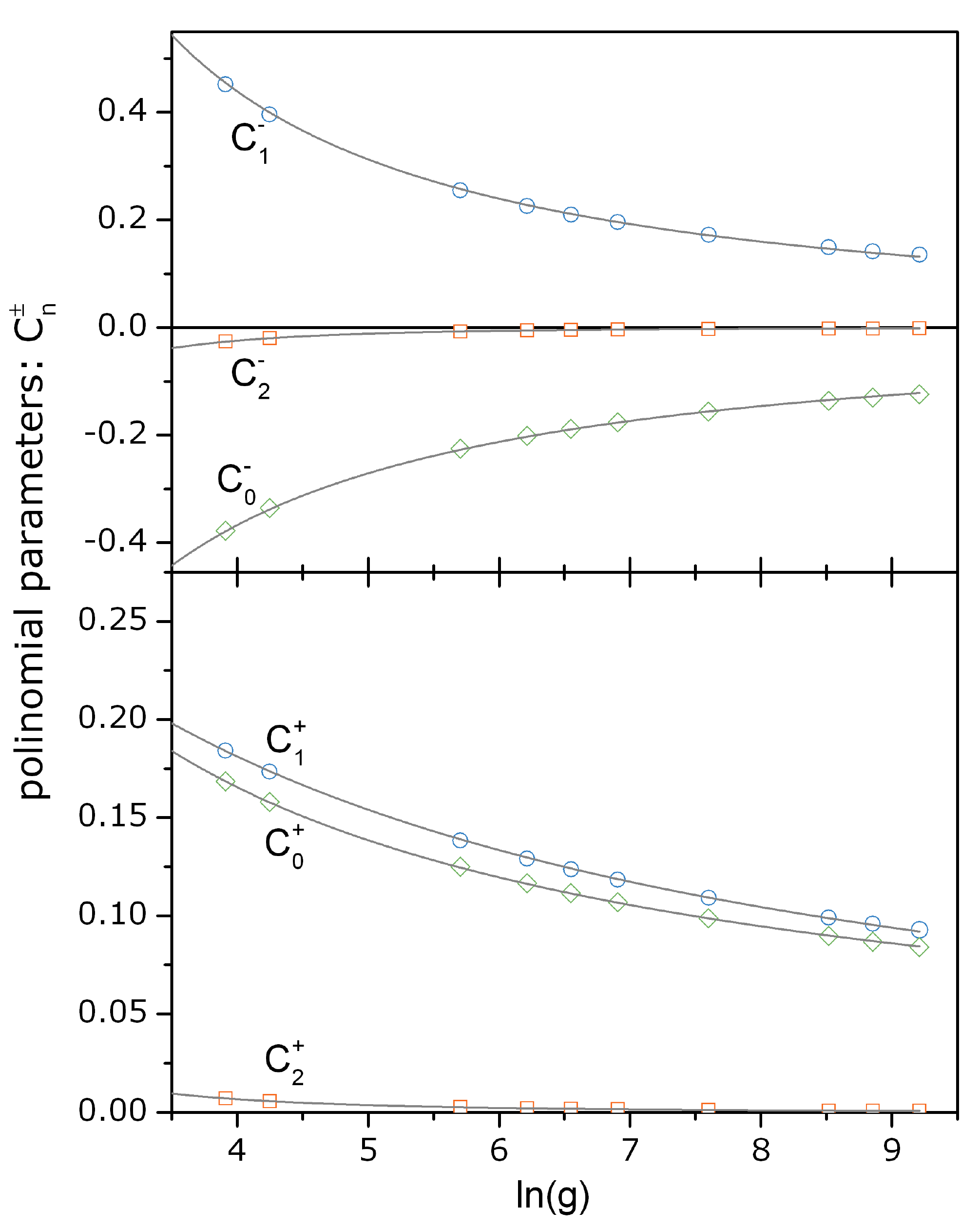}
	\caption{The hysteresis branches of the PD ($t_+$ and $t_-$) were fitted with a polynomial function, equation \ref{tmamee}. The panels (top for $t_-$ branch and bottom for $t_+$ branch) present these coefficients as a function of $\ln(g)$. The fits to these coefficients were made using equation \ref{coeffw}.} \label{para}
\end{figure}
\begin{equation}\label{coeffw}
    C_n^\pm=\left[\alpha_n^\pm\pm\beta_n^\pm\ln^{\delta_n^\pm}(g)\right]^{-1}.
\end{equation}

The parameters $\alpha_n^\pm$, $\beta_n^\pm$ and $\delta_n^\pm$ determining the PDs (for any value of $10^2\lesssim g \lesssim 10^4$) are listed in Table \ref{tab_par}. Note, in Figure \ref{para}, that for $g\gtrsim 250$ (i.e., $\ln(g)\gtrsim 5.5$) the quadratic term from equation \ref{tmamee} can be neglected.

\begin{table}
\caption{Parameters obtained  from the fit of the hysteresis branches of the PD ($t_+$ and $t_-$) with a polynomial function: equation \ref{tmamee}. These coefficients, as a function of $\ln(g)$, are presented on Figure \ref{para}. Fittings to these coefficients were made using equation \ref{coeffw}.\label{tab_par}}
\begin{tabular}{l|c|c|c}
 & \multicolumn{1}{l|}{$\;\;\;\alpha_n^\pm\;\;\;$} & \multicolumn{1}{l|}{$\;\;\;\beta_n^\pm\;\;\;$} & \multicolumn{1}{l}{$\;\;\;\delta_n^\pm\;\;\;$} \\ \hline
$C_0^+$ & 0.34(8) & 1.78(5) & 0.84(1) \\[0.5em]
$C_1^+$ & 2.21(8) & 0.67(4) & 1.15(3) \\[0.5em]
$C_2^+$ & -3.2(1) & 3.99(3) & 2.63(1) \\[0.5em] \hline
$C_0^-$ & 0.32(4) & -0.55(2) & 1.24(2) \\[0.5em]
$C_1^-$ & -0.52(4) & 0.47(2) & 1.28(2) \\[0.5em]
$C_2^-$ & -0.12(6) & -0.34(1) & 3.46(2)
\end{tabular}
\end{table}

\section{Application: Barocaloric effect}

This section explores the barocaloric effect (BCE). For this purpose, a benchmark material will be described in terms of structure and thermodynamic properties; as well as its corresponding entropy contributions. Finally, the results from the mean field model will be compared with experimental data.

\subsection{Entropy contributions}\label{entconre}

In order to obtain the entropy contributions, the HS molar fraction $n_{HS}=n_{HS}(T,P)$ must be determined, as described earlier in the manuscript. This fraction can be obtained from the virtual magnetization $m$ (that results from equation \ref{tmm1}) or from the self-consistent equation \ref{mself}, using the equilibrium effective field (equation \ref{heff3}). 

The entropy contributions for these materials have been previously addressed \cite{vallone2019giant,molnar2019molecular,gutlich1994thermal,fultz2010vibrational,von2021refrigeration}; and it is agreed that there are four main terms for the total entropy:
\begin{equation}
    S(T,P,B)=S_c(T,P)+S_s(T,P,B)+S_{l,o}(T,P)+S_{l,a}(T).
\end{equation}
These are, in order: configurational entropy, related with the spatial distribution of HS and LS molecules; the spin entropy and the last two account for lattice contributions (optical and acoustic phonons).

The barocaloric effect (BCE), on the other hand, depends on the entropy change due to the application of pressure. It can be written as:
\begin{equation}\label{bcefimf3ewd8}
    \Delta S(T,B,\Delta P)=
    \Delta S_c(T,\Delta P)+
    \Delta S_s(T,B,\Delta P)+
    \Delta S_{l,o}(T,\Delta P)
\end{equation}
This assumes that in the pressure range considered - only a few hundred bar - the acoustic entropy is pressure independent (in a first approximation), and thus does not contribute to barocaloric effect.

\subsubsection{HS-LS spatial distribution entropy term}

The entropy due to the spatial distribution of HS-LS molecules (configurational entropy) $S_{c}(T,P)$ can be obtained from the equilibrium free energy (equation \ref{Feqqq}), by using the standard thermodynamic relationship $S=-\partial F/\partial T$. Thus, we obtain for the configurational (molar) entropy:
\begin{equation}
    S_{c}(T,P)=R\left\{\ln\left[2\cosh\left(\beta\tilde{h}_{eff}\right)\right]-\beta\tilde{h}_{eff}m\right\}.
\end{equation}
This contribution is the same if written as:
\begin{align}\nonumber
    S_{c}(T,P)=-R&\{n_{HS}(T,P)\ln\left[n_{HS}(T,P)\right]\\
    &+\left[1-n_{HS}(T,P)\right]\ln\left[1-n_{HS}(T,P)\right]\}.
\end{align}

\subsubsection{Spin entropy term}
This contribution can be written as the spin (molar) entropy of the HS molecules (times the corresponding molar fraction) plus the spin (molar) entropy of the LS molecules (times the corresponding molar fraction), as follows:
\begin{align}\nonumber
    S_{s}(T,P,B)=&n_{HS}(T,P)S_s^{HS}(T,B)\\
    &+[1-n_{HS}(T,P)]S_s^{LS}(T,B).
\end{align}

The spin molar contribution for each state (either LS or HS), is given by:
\begin{align}
    S_{s}^i(T,B)=R\left\{\ln\left[Z(x_i)\right]-x_im_i\right\},
\end{align}
where
\begin{align}
Z(x_i)=\frac{\sinh(a_ix_i)}{\sinh(b_ix_i)}
\end{align}
is the spin partition function, $a_i=1+b_i$ and $b_i=1/2j_i$. For these equations, $j_i$ is the total angular momentum for each spin state. In addition, $x_i=\bar{g}_ij_i\beta \mu_BB$, where $\bar{g}_i$ is the Land\'{e} factor for each spin state and $B$ the applied magnetic field. Finally, $m_i=a_i\coth(a_ix_i)-b_i\coth(b_ix_i)$ is the real magnetization for each the spin state, given by the Brillouin function. For further details about this spin magnetization, see reference \cite{reis2013fundamentals}.

\subsubsection{Lattice entropy term I: Optical}

This contribution takes into account optical phonons \cite{fultz2010vibrational,molnar2019molecular}. Considering the metal at the center of an octahedron has $\nu$ modes of vibration with corresponding frequencies $\omega_\nu$, and each mode behaves as a quantum harmonic oscillator, the energy per octahedron can be written as: $\epsilon_i=\sum_\nu(n_\nu+1/2)\hbar\omega_\nu^i$, where $i$ stands for LS or HS state. The corresponding partition function is:
\begin{equation}
    Z_i(T)=\prod_\nu\frac{1}{2\sinh(\beta\hbar\omega_\nu^i/2)};
\end{equation}
with the (molar) entropy due to the optical phonons:
\begin{equation}\label{oen}
    S_{l,o}^i(T)=R\sum_\nu\left\{-\ln\left(1-e^{-\beta\hbar\omega_\nu^i}\right)+\frac{\beta\hbar\omega_\nu^i}{e^{\beta\hbar\omega_\nu^i}-1}\right\}.
\end{equation}

Considering HS and LS molecules, we can write the overall contribution due to the optical phonons as: equation \ref{oen} for molar fraction of molecules in the HS state plus the same for molecules in the LS state, as follows:
\begin{align}\nonumber
    S_{l,o}(T,P)=&n_{HS}(T,P)S_{l,o}^{HS}(T)\\
    &+[1-n_{HS}(T,P)]S_{l,o}^{LS}(T).
\end{align}

Molnar and co-workers\cite{molnar2019molecular} pointed out that low-frequency modes ($<400$ cm$^{-1}$ ) contribute significantly to this optical contribution; and the well known $3n-6=15$ modes of the coordination octahedron have a predominant contribution of 70\%. The conclusion is that these 15 modes are major contributors for the overall lattice entropy; however, other intramolecular modes also play a role and must be taken into account for the SCO phenomenon. A calculation and assignment of the possible modes has been performed via DFT and will be used later in section \ref{modesdft} for assessing how well this model matches the benchmark material.

\subsubsection{Lattice entropy term II: Acoustic}

This contribution takes into account the acoustic phonons. In this context we use the standard Debye term, where the specific heat $c=c(T)$ is considered in its standard form (see references \cite{ashcroft1976solid} and  \cite{reis2020magnetocaloric}):
\begin{equation}
    c(T)=9R\left(\frac{T}{T_D}\right)^3\int_0^{T_D/T}\frac{x^4e^x}{(e^x-1)^2}dx.
\end{equation}
From the above result, the molar entropy can be obtained from the standard thermodynamic equations: $S(T)=\int_0^T(c/T)\,dT$. Note the acoustic contribution $S_{l,a}(T)$ only depends on the temperature (in the first approximation) and, consequently, the total entropy change due to external excitation (magnetic field, pressure, etc), does not depend on this term.

\subsection{Connections with experimental results}\label{expff}

\subsubsection{Material description}

As a benchmark material, we will use the molecular compound [FeL$_2$][BF$_4$]$_2$, where L is [L=2,6-di(pyrazol-1-yl)pyridine] (BPP). This material was reported to exhibit a giant barocaloric effect associated with the SCO transition \cite{vallone2019giant}. The present discussion applies the mean field model described above to this material and assesses how suitable it is in predicting its BCE performance, as well as how the different contributions to the entropy change are partitioned.

\begin{figure}
\begin{center}
\subfigure[]{
\includegraphics[width=1.1\columnwidth,keepaspectratio]{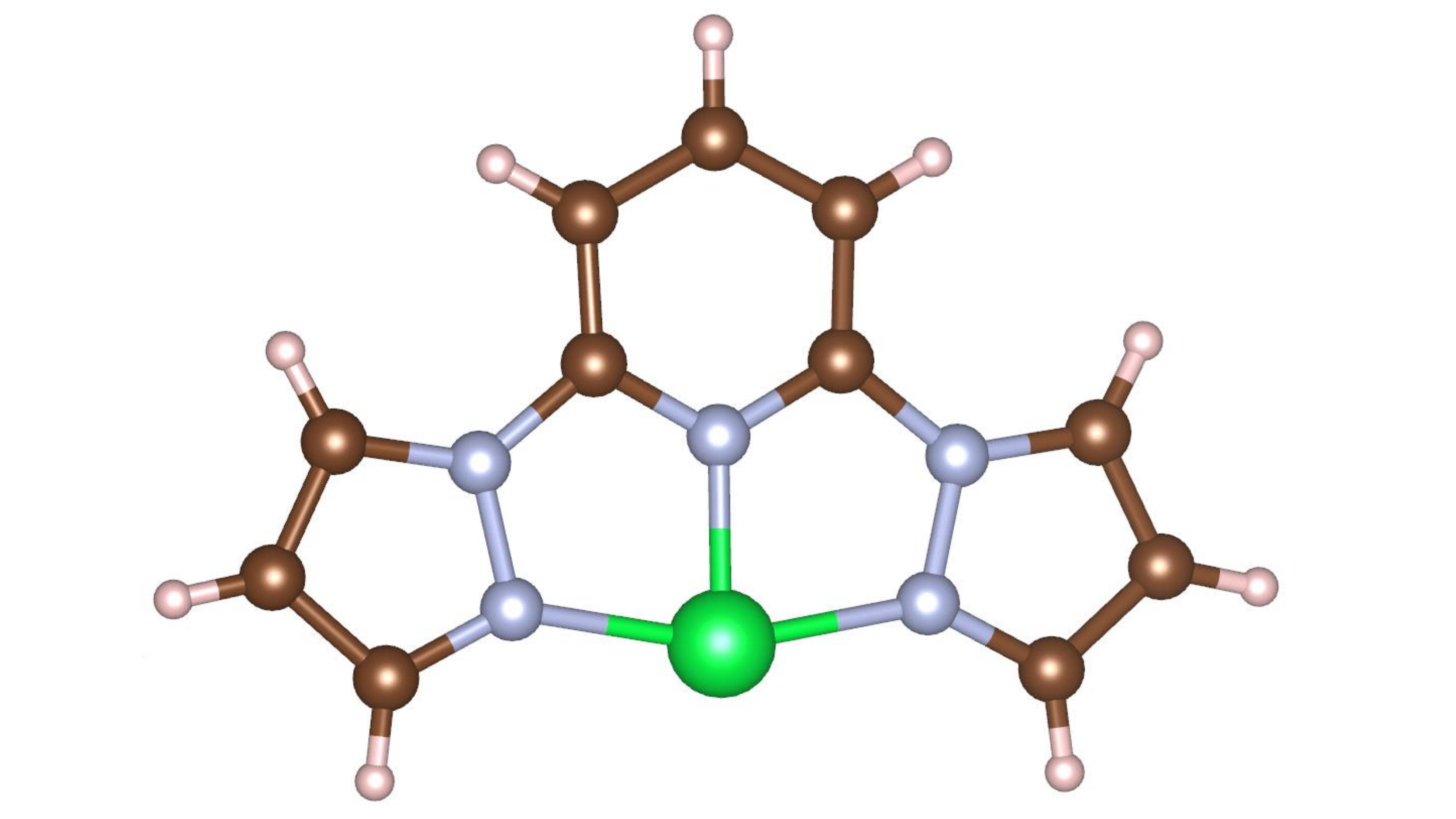}
}\hfill
\subfigure[]{
\includegraphics[width=0.5\columnwidth,keepaspectratio]{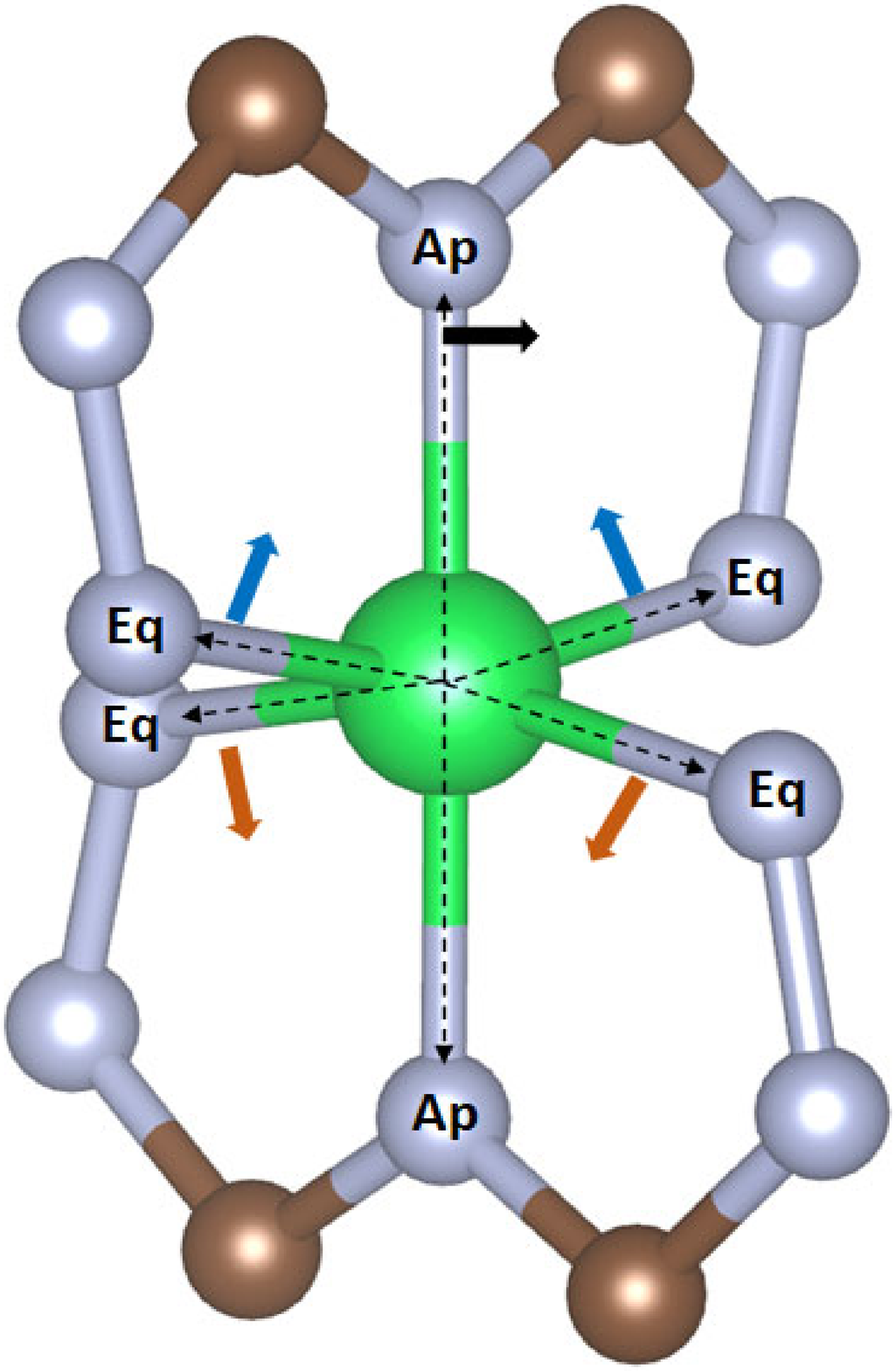}
}
\end{center}
\caption{(a) Structural representation of 1-BPP ligand and its coordination with Fe. (b) Detail of the FeN$_{6}$ octahedral coordination. Arrows highlight the local octhaedral average distortion when transitioning from the LS to HS state. Color codes:  Fe-Green, N–Gray, C-Brown and H-White.\label{bpp}}
\end{figure}

The BPP ligand consists of a planar arrangement of two pyrazoles, surrounding a central pyridine, as shown in Figure \ref{bpp}(a). Three nitrogen atoms of the ligand coordinate with the central iron ion. In the crystal form, two BPP ligands arranged perpendicularly about the iron form a slightly distorted FeN$_6$ octahedron, as seen in Figure \ref{bpp}(b). In this coordination, there are two opposing FeN$_{pyridine}$, while four FeN$_{pyrrole}$ are slightly staggered above and below the equatorial plane (see arrows). This FeN$_6$ octahedra presents, at all temperatures, a slightly distorted $D_{2d}$ symmetry.

The Fe(BPP)$_2$ complex has an overall +2 charge that must be balanced with counter ions. In the present compound the charge balance is achieved via two BF$_4$ anions. This compound has been shown to undergo a LS to HS transition centered at 259 K, with a narrow (3 K) hysteresis \cite{holland2001unusual}. From the structural point of view, the LS to HS transition results in an increased Fe coordination volume, consistent with the occupancy of the most outward $d$ orbitals.

The source data for this benchmark was gathered in part, from in-situ high pressure neutron powder diffraction data collected at the SNAP beamline at the SNS\cite{vallone2019giant}, to determine the spin transitions temperatures at each pressure, along with the corresponding unit cell volume. High pressure magnetization and calorimetry are also reported in reference \onlinecite{vallone2019giant}, and references within.    
Table \ref{tab_par} summarizes the main structural parameters, such as average bond length, octahedral and cell volumes and angles, for LS and HS states. It is noteworthy that the LS to HS transition corresponds to c.a. 30 \% increase in the Fe first neighbors coordination sphere ($V_O$), but only 5\% in the unit cell ($V_C$), regardless of the applied pressure. However, in absolute terms, the volume change in the unit cell is much larger. These values are in accordance with the literature\cite{molnar2019molecular}, where typical volume changes for octahedra are close to 25\%. As mentioned by Molnar and co-workers\cite{molnar2019molecular}, these structural changes are responsible for changes on the vibrational spectra, measured by Raman scattering, IR absorption spectroscopies and DFT calculations. This fact will be discussed in detail further below. This provides a structural illustration of how the SCO transition merely \emph{triggers} the caloric effect, while the structure as a whole is responsible for the caloric performance. \\
\begin{table}[]
\caption{Main structural parameters for [FeL$_2$][BF$_4$]$_2$, where L is a [L=2,6-di(pyrazol-1-yl)pyridine] (BPP). $V_O$: octhaedral volume. $V_C$: cell volume. The data was obtained from neutron powder data. Refinment of the octahedral dimensions at high pressure was not possible. \label{tab_par}}
\begin{tabular}{c|c|c|c|c}
& \multicolumn{2}{c|}{LS}& \multicolumn{2}{c}{HS}\\ \hline\hline
Average bond length (\AA)& \multicolumn{2}{c|}{1.9580}     & \multicolumn{2}{c}{2.1785}\\
$N_{ap}-Fe-N_{ap}$ & \multicolumn{2}{c|}{178$^\circ$}        & \multicolumn{2}{c}{173$^\circ$}\\
$N_{eq}-Fe-N_{eq}$ & \multicolumn{2}{c|}{160$^\circ$}        & \multicolumn{2}{c}{146$^\circ$}\\ \hline\hline
\multicolumn{1}{c|}{\begin{tabular}[c]{@{}c@{}}Pressure\\ (MPa)\end{tabular}} 
   & \multicolumn{1}{c|}{\begin{tabular}[c]{@{}c@{}}$V_O$\\ (\AA$^3)$\end{tabular}} 
   & \multicolumn{1}{c|}{\begin{tabular}[c]{@{}c@{}}$V_C$\\ (\AA$^3)$\end{tabular}} 
   & \multicolumn{1}{c|}{\begin{tabular}[c]{@{}c@{}}$V_O$\\ (\AA$^3)$\end{tabular}} 
   & \multicolumn{1}{c}{\begin{tabular}[c]{@{}c@{}}$V_C$\\ (\AA$^3)$\end{tabular}} 
   \\\hline
  ambient  & 9.7 & 1331 & 12.6 & 1365                \\
  20  & - &1329& - & 1360 \\
\end{tabular}
\end{table}

\subsubsection{Modes of vibrations: DFT}\label{modesdft}


The vibration spectra of the Fe(BPP)$_3$ clusters in the LS and HS configuration, were calculated using Gaussian 09 \cite{frisch200901}, using starting configurations extracted from single crystal diffraction \cite{holland2001unusual}. Following the method reported by Cirera et al. \cite{cirera2018benchmarking}, TPSSh functional was adopted, with QZPV basis set on the metal center and TZV basis for all the other elements. Structure optimization and vibrational analysis were performed, and the calculation results agree with those reported in reference \onlinecite{cirera2018benchmarking}. The normal modes from the vibrational analysis were then used to calculate the partial phonon density of states, as well as to visualize the polarization vectors with Jmol \cite{jmol}.

\begin{figure}
	\centering
	\includegraphics[width=\columnwidth,keepaspectratio]{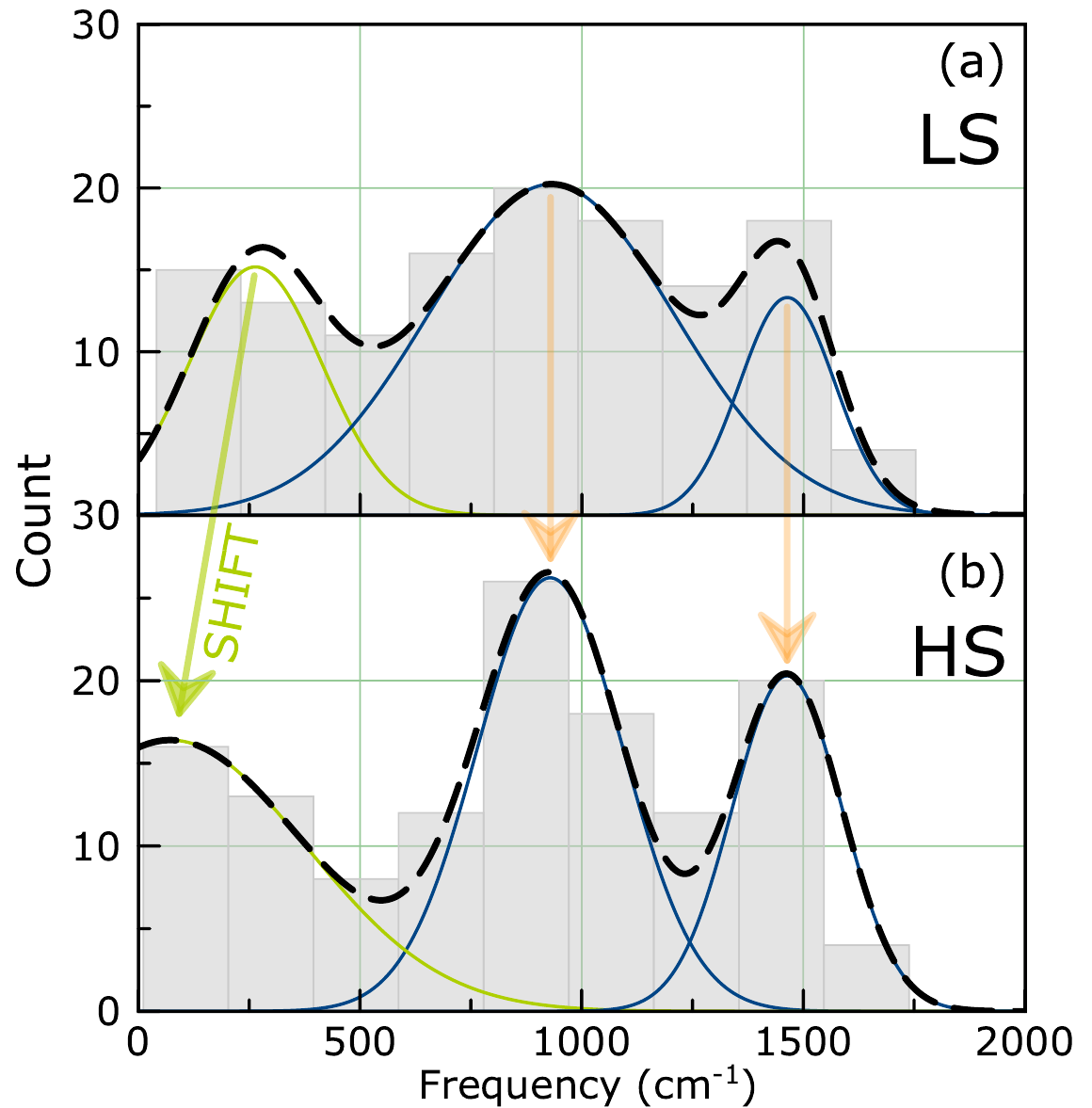}
	\caption{Frequency histogram for all possible modes of vibration, considering (a) LS and (b) HS states. \label{histo}}
\end{figure}

Figure \ref{histo} represents a frequency histogram for all possible modes of vibration, considering the LS state, panel (a), and HS state, panel (b). Both states show a clear tri-modal behavior, corresponding to predominant types of vibration. It is clear from these results that the two groups corresponding to higher frequency modes -- the one at high frequencies (close to 1450 cm$^{-1}$) and the one at intermediate frequencies (close to 900 cm$^{-1}$) -- are insensitive to the SCO transition. In contrast, the modes at low frequencies exhibit a remarkable shift in frequency across the SCO transition, changing, on average, from 264 cm$^{-1}$ (LS state) down to 72 cm$^{-1}$ (HS state). In addition, the \textit{mode} of the Gaussian fits indicates that about 15 modes are involved in these vibrational changes. This observation is in agreement with Molnar and co-workers\cite{molnar2019molecular}, which stated that the low-frequency modes ($<400$ cm$^{-1}$ ) are the most important; and the 15 modes of vibration of the coordination octahedron would have a predominant contribution. These results are also supported by references \onlinecite{von2020large,ribeiro2019influence}. Finally, we note that the normal mode calculation of the isolated molecules does not take anharmonicity into consideration. However, we believe the effects are secondary compared to the other approximations used in the model, and the ``shift'' in frequency between LS and HS should not be affected.

\subsubsection{Barocaloric potentials}

In the work by Vallone and co-workers, in addition to the structural information, it  also reported experimental thermodynamic properties measured \emph{in-situ} under pressure, such as magnetization and calorimetry. From the latter, the absolute entropy at ambient conditions and under pressure (43 MPa) were obtained. Based on these results, the present work steps forward and evaluated the experimental barocaloric entropy change.
\begin{figure}
	\centering
	\includegraphics[width=\columnwidth,keepaspectratio]{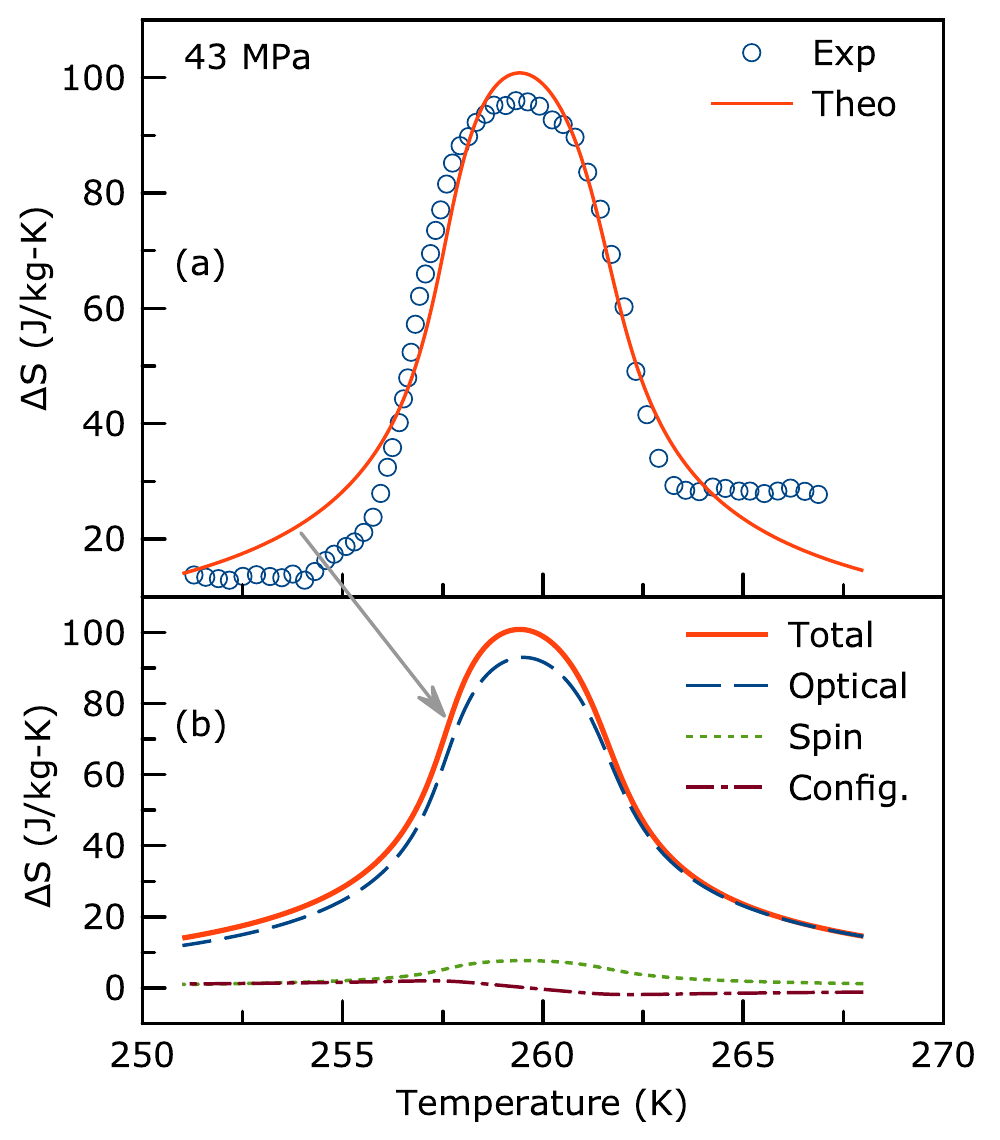}
	\caption{(a) Experimental and theoretical barocaloric entropy change. Experimental data are from reference \onlinecite{vallone2019giant}. Theoretical model is the preset work. (b) Entropy change contributions to the total entropy change, namely: optical phonons (responsible for 92 \% of the entropy change), spin and spatial distribution of HS-LS molecules. \label{fitting}}
\end{figure}

The fit of the theoretical model (see subsection \ref{entconre}) to the experimental data is presented in Figure \ref{fitting}(a). The \textit{modal} number of modes ($\bar{\nu}=15$) and the average frequency for these modes in the LS ($\bar{\omega}^{LS}_{\bar{\nu}}=264$ cm$^{-1}$) and HS ($\bar{\omega}^{HS}_{\bar{\nu}}=72$ cm$^{-1}$) states were fixed inputs to the model derived from the DFT calculation (see subsection \ref{modesdft}). The volume change $\delta V=35$ \AA$^3$ across the SCO transition under 43 MPa was obtained from reference \onlinecite{vallone2019giant}. The three free parameters allowed to vary were: the degeneracy ratio $g=862$ (see equation \ref{Deff}), the ligand-field energy $\Delta_0=1741$ K and the characteristic temperature $T_0=236$ K (see equation \ref{t0frec3}). It can be seen that this mean field approach does result in a very satisfactory match with experiment.

Figure \ref{fitting}(b) presents the partition of the entropy change in terms of the contributions discussed in subsection \ref{entconre}. There are three contributions to the total entropy change: lattice (optical phonons), spin and configurational. From these, optical phonon contribution is responsible for 92\% of the total entropy change, while the spin and configurational account for the remainder 8\%. As represented on Figure \ref{histo}, those $\bar{\nu}=15$ modes at low frequency (from $\bar{\omega}^{LS}_{\bar{\nu}}=264$ cm$^{-1}$ to $\bar{\omega}^{HS}_{\bar{\nu}}=72$ cm$^{-1}$) are responsible for this effect.

\section{Summary and Conclusions}

The mean field model for SCO transition, based on lattice elasticity, HS-LS molecule state and applied pressure, was proposed. From this model, the high spin molar fraction was then obtained as a function of temperature and pressure, where first or second order transitions could be induced, depending on the thermodynamic state of the sample. From these results, a comprehensive phase diagram can be constructed. In order to validate this approach we applied this strategy on a SCO-driven giant barocaloric system experimentally studied  \cite{vallone2019giant}. Several entropy contributions were considered, such as HS-LS spatial distribution of molecules, spin and lattice, with the last one containing optical and acoustic phonons. In addition, we performed DFT calculations to derive the vibration modes and corresponding frequencies. Fitting of the theoretical barocaloric entropy change to the experimental data produces a satisfactory matching. This fitting procedure allows for only three free parameters: the degeneracy ratio $g$, the ligand-field energy $\Delta_0$ and the characteristic temperature $T_0$. Finally, as a conclusion, it is found that the optical phonons are responsible for 92\% of the total barocaloric entropy change. From our DFT calculations, it is clear that among the three main groups of frequencies, the one at lower frequency, below 400 cm$^{-1}$, shifts from 264 cm$^{-1}$ down to 72 cm$^{-1}$, these are associated with Fe-N participating modes is responsible for the BCO. This is also in agreement with earlier observations by Molnar\cite{molnar2019molecular}. Given the recent interest in caloric effects, this model may present an accessible way to screen other potential systems that rely on SCO as a trigger for the BCE as well as other SCO applications.

\begin{acknowledgments}

A portion of this research used resources at the Spallation Neutron Source, a DOE Office of Science User Facility operated by the Oak Ridge National Laboratory. The computing resources for DFT calculations were made available through the VirtuES project, funded by Laboratory Directed Research and Development program and Compute and Data Environment for Science (CADES) at ORNL. MSR thanks CNPq and FAPERJ.


\end{acknowledgments}

\section{Author's contributions}
MSR: research conception, thermodynamic model, application and analysis; YQC: DFT results and analysis; AMS: material description and analysis. All the authors contributed to the article writing. 
\section{Data availability}
The data that support the findings of this study are available from the corresponding author upon reasonable request.


\begin{thebibliography}{19}%
\makeatletter
\providecommand \@ifxundefined [1]{%
 \@ifx{#1\undefined}
}%
\providecommand \@ifnum [1]{%
 \ifnum #1\expandafter \@firstoftwo
 \else \expandafter \@secondoftwo
 \fi
}%
\providecommand \@ifx [1]{%
 \ifx #1\expandafter \@firstoftwo
 \else \expandafter \@secondoftwo
 \fi
}%
\providecommand \natexlab [1]{#1}%
\providecommand \enquote  [1]{``#1''}%
\providecommand \bibnamefont  [1]{#1}%
\providecommand \bibfnamefont [1]{#1}%
\providecommand \citenamefont [1]{#1}%
\providecommand \href@noop [0]{\@secondoftwo}%
\providecommand \href [0]{\begingroup \@sanitize@url \@href}%
\providecommand \@href[1]{\@@startlink{#1}\@@href}%
\providecommand \@@href[1]{\endgroup#1\@@endlink}%
\providecommand \@sanitize@url [0]{\catcode `\\12\catcode `\$12\catcode
  `\&12\catcode `\#12\catcode `\^12\catcode `\_12\catcode `\%12\relax}%
\providecommand \@@startlink[1]{}%
\providecommand \@@endlink[0]{}%
\providecommand \url  [0]{\begingroup\@sanitize@url \@url }%
\providecommand \@url [1]{\endgroup\@href {#1}{\urlprefix }}%
\providecommand \urlprefix  [0]{URL }%
\providecommand \Eprint [0]{\href }%
\providecommand \doibase [0]{http://dx.doi.org/}%
\providecommand \selectlanguage [0]{\@gobble}%
\providecommand \bibinfo  [0]{\@secondoftwo}%
\providecommand \bibfield  [0]{\@secondoftwo}%
\providecommand \translation [1]{[#1]}%
\providecommand \BibitemOpen [0]{}%
\providecommand \bibitemStop [0]{}%
\providecommand \bibitemNoStop [0]{.\EOS\space}%
\providecommand \EOS [0]{\spacefactor3000\relax}%
\providecommand \BibitemShut  [1]{\csname bibitem#1\endcsname}%
\let\auto@bib@innerbib\@empty
\bibitem [{\citenamefont {Coronado}(2020)}]{coronado2020molecular}%
  \BibitemOpen
  \bibfield  {author} {\bibinfo {author} {\bibfnamefont {E.}~\bibnamefont
  {Coronado}},\ }\href@noop {} {\bibfield  {journal} {\bibinfo  {journal}
  {Nature Reviews Materials}\ }\textbf {\bibinfo {volume} {5}},\ \bibinfo
  {pages} {87} (\bibinfo {year} {2020})}\BibitemShut {NoStop}%
\bibitem [{\citenamefont {Sandeman}(2016)}]{sandeman2016research}%
  \BibitemOpen
  \bibfield  {author} {\bibinfo {author} {\bibfnamefont {K.~G.}\ \bibnamefont
  {Sandeman}},\ }\href@noop {} {\bibfield  {journal} {\bibinfo  {journal} {APL
  Materials}\ }\textbf {\bibinfo {volume} {4}},\ \bibinfo {pages} {111102}
  (\bibinfo {year} {2016})}\BibitemShut {NoStop}%
\bibitem [{\citenamefont {Vallone}\ \emph {et~al.}(2019)\citenamefont
  {Vallone}, \citenamefont {Tantillo}, \citenamefont {Dos~Santos},
  \citenamefont {Molaison}, \citenamefont {Kulmaczewski}, \citenamefont
  {Chapoy}, \citenamefont {Ahmadi}, \citenamefont {Halcrow},\ and\
  \citenamefont {Sandeman}}]{vallone2019giant}%
  \BibitemOpen
  \bibfield  {author} {\bibinfo {author} {\bibfnamefont {S.~P.}\ \bibnamefont
  {Vallone}}, \bibinfo {author} {\bibfnamefont {A.~N.}\ \bibnamefont
  {Tantillo}}, \bibinfo {author} {\bibfnamefont {A.~M.}\ \bibnamefont
  {Dos~Santos}}, \bibinfo {author} {\bibfnamefont {J.~J.}\ \bibnamefont
  {Molaison}}, \bibinfo {author} {\bibfnamefont {R.}~\bibnamefont
  {Kulmaczewski}}, \bibinfo {author} {\bibfnamefont {A.}~\bibnamefont
  {Chapoy}}, \bibinfo {author} {\bibfnamefont {P.}~\bibnamefont {Ahmadi}},
  \bibinfo {author} {\bibfnamefont {M.~A.}\ \bibnamefont {Halcrow}}, \ and\
  \bibinfo {author} {\bibfnamefont {K.~G.}\ \bibnamefont {Sandeman}},\
  }\href@noop {} {\bibfield  {journal} {\bibinfo  {journal} {Advanced
  Materials}\ }\textbf {\bibinfo {volume} {31}},\ \bibinfo {pages} {1807334}
  (\bibinfo {year} {2019})}\BibitemShut {NoStop}%
\bibitem [{\citenamefont {Ribeiro}\ \emph {et~al.}(2019)\citenamefont
  {Ribeiro}, \citenamefont {Alho}, \citenamefont {Ribas}, \citenamefont
  {N{\'o}brega}, \citenamefont {de~Sousa},\ and\ \citenamefont {von
  Ranke}}]{ribeiro2019influence}%
  \BibitemOpen
  \bibfield  {author} {\bibinfo {author} {\bibfnamefont {P.}~\bibnamefont
  {Ribeiro}}, \bibinfo {author} {\bibfnamefont {B.}~\bibnamefont {Alho}},
  \bibinfo {author} {\bibfnamefont {R.}~\bibnamefont {Ribas}}, \bibinfo
  {author} {\bibfnamefont {E.}~\bibnamefont {N{\'o}brega}}, \bibinfo {author}
  {\bibfnamefont {V.}~\bibnamefont {de~Sousa}}, \ and\ \bibinfo {author}
  {\bibfnamefont {P.}~\bibnamefont {von Ranke}},\ }\href@noop {} {\bibfield
  {journal} {\bibinfo  {journal} {Journal of Magnetism and Magnetic Materials}\
  }\textbf {\bibinfo {volume} {489}},\ \bibinfo {pages} {165340} (\bibinfo
  {year} {2019})}\BibitemShut {NoStop}%
\bibitem [{\citenamefont {von Ranke}\ \emph {et~al.}(2021)\citenamefont {von
  Ranke}, \citenamefont {Alho}, \citenamefont {da~Silva}, \citenamefont
  {Ribas}, \citenamefont {Nobrega}, \citenamefont {de~Sousa}, \citenamefont
  {Carvalho},\ and\ \citenamefont {Ribeiro}}]{von2021refrigeration}%
  \BibitemOpen
  \bibfield  {author} {\bibinfo {author} {\bibfnamefont {P.~J.}\ \bibnamefont
  {von Ranke}}, \bibinfo {author} {\bibfnamefont {B.~P.}\ \bibnamefont {Alho}},
  \bibinfo {author} {\bibfnamefont {P.~H.}\ \bibnamefont {da~Silva}}, \bibinfo
  {author} {\bibfnamefont {R.~M.}\ \bibnamefont {Ribas}}, \bibinfo {author}
  {\bibfnamefont {E.~P.}\ \bibnamefont {Nobrega}}, \bibinfo {author}
  {\bibfnamefont {V.~S.}\ \bibnamefont {de~Sousa}}, \bibinfo {author}
  {\bibfnamefont {A.~M.}\ \bibnamefont {Carvalho}}, \ and\ \bibinfo {author}
  {\bibfnamefont {P.~O.}\ \bibnamefont {Ribeiro}},\ }\href@noop {} {\bibfield
  {journal} {\bibinfo  {journal} {physica status solidi (b)}\ }\textbf
  {\bibinfo {volume} {258}},\ \bibinfo {pages} {2100108} (\bibinfo {year}
  {2021})}\BibitemShut {NoStop}%
\bibitem [{\citenamefont {Von~Ranke}\ \emph {et~al.}(2020)\citenamefont
  {Von~Ranke}, \citenamefont {Alho}, \citenamefont {da~Silva}, \citenamefont
  {Ribas}, \citenamefont {Nobrega}, \citenamefont {De~Sousa}, \citenamefont
  {Cola{\c{c}}o}, \citenamefont {Marques}, \citenamefont {Reis}, \citenamefont
  {Scaldini} \emph {et~al.}}]{von2020large}%
  \BibitemOpen
  \bibfield  {author} {\bibinfo {author} {\bibfnamefont {P.}~\bibnamefont
  {Von~Ranke}}, \bibinfo {author} {\bibfnamefont {B.}~\bibnamefont {Alho}},
  \bibinfo {author} {\bibfnamefont {P.}~\bibnamefont {da~Silva}}, \bibinfo
  {author} {\bibfnamefont {R.}~\bibnamefont {Ribas}}, \bibinfo {author}
  {\bibfnamefont {E.}~\bibnamefont {Nobrega}}, \bibinfo {author} {\bibfnamefont
  {V.}~\bibnamefont {De~Sousa}}, \bibinfo {author} {\bibfnamefont
  {M.}~\bibnamefont {Cola{\c{c}}o}}, \bibinfo {author} {\bibfnamefont {L.~F.}\
  \bibnamefont {Marques}}, \bibinfo {author} {\bibfnamefont {M.}~\bibnamefont
  {Reis}}, \bibinfo {author} {\bibfnamefont {F.}~\bibnamefont {Scaldini}},
  \emph {et~al.},\ }\href@noop {} {\bibfield  {journal} {\bibinfo  {journal}
  {Journal of Applied Physics}\ }\textbf {\bibinfo {volume} {127}},\ \bibinfo
  {pages} {165104} (\bibinfo {year} {2020})}\BibitemShut {NoStop}%
\bibitem [{\citenamefont {Babilotte}\ and\ \citenamefont
  {Boukheddaden}(2020)}]{babilotte2020theoretical}%
  \BibitemOpen
  \bibfield  {author} {\bibinfo {author} {\bibfnamefont {K.}~\bibnamefont
  {Babilotte}}\ and\ \bibinfo {author} {\bibfnamefont {K.}~\bibnamefont
  {Boukheddaden}},\ }\href@noop {} {\bibfield  {journal} {\bibinfo  {journal}
  {Physical Review B}\ }\textbf {\bibinfo {volume} {101}},\ \bibinfo {pages}
  {174113} (\bibinfo {year} {2020})}\BibitemShut {NoStop}%
\bibitem [{\citenamefont {von Ranke}(2017)}]{von2017microscopic}%
  \BibitemOpen
  \bibfield  {author} {\bibinfo {author} {\bibfnamefont {P.}~\bibnamefont {von
  Ranke}},\ }\href@noop {} {\bibfield  {journal} {\bibinfo  {journal} {Applied
  Physics Letters}\ }\textbf {\bibinfo {volume} {110}},\ \bibinfo {pages}
  {181909} (\bibinfo {year} {2017})}\BibitemShut {NoStop}%
\bibitem [{\citenamefont {von Ranke}\ \emph {et~al.}(2022)\citenamefont {von
  Ranke}, \citenamefont {Alho}, \citenamefont {Ribas}, \citenamefont {Nobrega},
  \citenamefont {de~Sousa},\ and\ \citenamefont
  {Ribeiro}}]{von2022theoretical}%
  \BibitemOpen
  \bibfield  {author} {\bibinfo {author} {\bibfnamefont {P.}~\bibnamefont {von
  Ranke}}, \bibinfo {author} {\bibfnamefont {B.}~\bibnamefont {Alho}}, \bibinfo
  {author} {\bibfnamefont {R.}~\bibnamefont {Ribas}}, \bibinfo {author}
  {\bibfnamefont {E.}~\bibnamefont {Nobrega}}, \bibinfo {author} {\bibfnamefont
  {V.}~\bibnamefont {de~Sousa}}, \ and\ \bibinfo {author} {\bibfnamefont
  {P.}~\bibnamefont {Ribeiro}},\ }\href@noop {} {\bibfield  {journal} {\bibinfo
   {journal} {Journal of Magnetism and Magnetic Materials}\ }\textbf {\bibinfo
  {volume} {564}},\ \bibinfo {pages} {170121} (\bibinfo {year}
  {2022})}\BibitemShut {NoStop}%
\bibitem [{\citenamefont {Moln{\'a}r}\ \emph {et~al.}(2019)\citenamefont
  {Moln{\'a}r}, \citenamefont {Mikolasek}, \citenamefont {Ridier},
  \citenamefont {Fahs}, \citenamefont {Nicolazzi},\ and\ \citenamefont
  {Bousseksou}}]{molnar2019molecular}%
  \BibitemOpen
  \bibfield  {author} {\bibinfo {author} {\bibfnamefont {G.}~\bibnamefont
  {Moln{\'a}r}}, \bibinfo {author} {\bibfnamefont {M.}~\bibnamefont
  {Mikolasek}}, \bibinfo {author} {\bibfnamefont {K.}~\bibnamefont {Ridier}},
  \bibinfo {author} {\bibfnamefont {A.}~\bibnamefont {Fahs}}, \bibinfo {author}
  {\bibfnamefont {W.}~\bibnamefont {Nicolazzi}}, \ and\ \bibinfo {author}
  {\bibfnamefont {A.}~\bibnamefont {Bousseksou}},\ }\href@noop {} {\bibfield
  {journal} {\bibinfo  {journal} {Annalen der Physik}\ }\textbf {\bibinfo
  {volume} {531}},\ \bibinfo {pages} {1900076} (\bibinfo {year}
  {2019})}\BibitemShut {NoStop}%
\bibitem [{\citenamefont {G{\"u}tlich}\ \emph {et~al.}(1994)\citenamefont
  {G{\"u}tlich}, \citenamefont {Hauser},\ and\ \citenamefont
  {Spiering}}]{gutlich1994thermal}%
  \BibitemOpen
  \bibfield  {author} {\bibinfo {author} {\bibfnamefont {P.}~\bibnamefont
  {G{\"u}tlich}}, \bibinfo {author} {\bibfnamefont {A.}~\bibnamefont {Hauser}},
  \ and\ \bibinfo {author} {\bibfnamefont {H.}~\bibnamefont {Spiering}},\
  }\href@noop {} {\bibfield  {journal} {\bibinfo  {journal} {Angewandte Chemie
  International Edition in English}\ }\textbf {\bibinfo {volume} {33}},\
  \bibinfo {pages} {2024} (\bibinfo {year} {1994})}\BibitemShut {NoStop}%
\bibitem [{\citenamefont {Fultz}(2010)}]{fultz2010vibrational}%
  \BibitemOpen
  \bibfield  {author} {\bibinfo {author} {\bibfnamefont {B.}~\bibnamefont
  {Fultz}},\ }\href@noop {} {\bibfield  {journal} {\bibinfo  {journal}
  {Progress in Materials Science}\ }\textbf {\bibinfo {volume} {55}},\ \bibinfo
  {pages} {247} (\bibinfo {year} {2010})}\BibitemShut {NoStop}%
\bibitem [{\citenamefont {Reis}(2013)}]{reis2013fundamentals}%
  \BibitemOpen
  \bibfield  {author} {\bibinfo {author} {\bibfnamefont {M.}~\bibnamefont
  {Reis}},\ }\href@noop {} {\emph {\bibinfo {title} {Fundamentals of
  magnetism}}}\ (\bibinfo  {publisher} {Elsevier},\ \bibinfo {year}
  {2013})\BibitemShut {NoStop}%
\bibitem [{\citenamefont {Ashcroft}\ \emph {et~al.}(1976)\citenamefont
  {Ashcroft}, \citenamefont {Mermin} \emph {et~al.}}]{ashcroft1976solid}%
  \BibitemOpen
  \bibfield  {author} {\bibinfo {author} {\bibfnamefont {N.~W.}\ \bibnamefont
  {Ashcroft}}, \bibinfo {author} {\bibfnamefont {N.~D.}\ \bibnamefont
  {Mermin}},  \emph {et~al.},\ }\href@noop {} {\enquote {\bibinfo {title}
  {Solid state physics},}\ } (\bibinfo {year} {1976})\BibitemShut {NoStop}%
\bibitem [{\citenamefont {Reis}(2020)}]{reis2020magnetocaloric}%
  \BibitemOpen
  \bibfield  {author} {\bibinfo {author} {\bibfnamefont {M.~S.}\ \bibnamefont
  {Reis}},\ }\href@noop {} {\bibfield  {journal} {\bibinfo  {journal}
  {Coordination Chemistry Reviews}\ }\textbf {\bibinfo {volume} {417}},\
  \bibinfo {pages} {213357} (\bibinfo {year} {2020})}\BibitemShut {NoStop}%
\bibitem [{\citenamefont {Holland}\ \emph {et~al.}(2001)\citenamefont
  {Holland}, \citenamefont {McAllister}, \citenamefont {Lu}, \citenamefont
  {Kilner}, \citenamefont {Thornton-Pett},\ and\ \citenamefont
  {Halcrow}}]{holland2001unusual}%
  \BibitemOpen
  \bibfield  {author} {\bibinfo {author} {\bibfnamefont {J.~M.}\ \bibnamefont
  {Holland}}, \bibinfo {author} {\bibfnamefont {J.~A.}\ \bibnamefont
  {McAllister}}, \bibinfo {author} {\bibfnamefont {Z.}~\bibnamefont {Lu}},
  \bibinfo {author} {\bibfnamefont {C.~A.}\ \bibnamefont {Kilner}}, \bibinfo
  {author} {\bibfnamefont {M.}~\bibnamefont {Thornton-Pett}}, \ and\ \bibinfo
  {author} {\bibfnamefont {M.~A.}\ \bibnamefont {Halcrow}},\ }\href@noop {}
  {\bibfield  {journal} {\bibinfo  {journal} {Chemical Communications}\ ,\
  \bibinfo {pages} {577}} (\bibinfo {year} {2001})}\BibitemShut {NoStop}%
\bibitem [{\citenamefont {Frisch}\ \emph {et~al.}(2009)\citenamefont {Frisch},
  \citenamefont {Trucks}, \citenamefont {Schlegel}, \citenamefont {Scuseria},
  \citenamefont {Robb}, \citenamefont {Cheeseman}, \citenamefont {Scalmani},
  \citenamefont {Barone}, \citenamefont {Mennucci}, \citenamefont {Petersson}
  \emph {et~al.}}]{frisch200901}%
  \BibitemOpen
  \bibfield  {author} {\bibinfo {author} {\bibfnamefont {M.}~\bibnamefont
  {Frisch}}, \bibinfo {author} {\bibfnamefont {G.}~\bibnamefont {Trucks}},
  \bibinfo {author} {\bibfnamefont {H.}~\bibnamefont {Schlegel}}, \bibinfo
  {author} {\bibfnamefont {G.}~\bibnamefont {Scuseria}}, \bibinfo {author}
  {\bibfnamefont {M.}~\bibnamefont {Robb}}, \bibinfo {author} {\bibfnamefont
  {J.}~\bibnamefont {Cheeseman}}, \bibinfo {author} {\bibfnamefont
  {G.}~\bibnamefont {Scalmani}}, \bibinfo {author} {\bibfnamefont
  {V.}~\bibnamefont {Barone}}, \bibinfo {author} {\bibfnamefont
  {B.}~\bibnamefont {Mennucci}}, \bibinfo {author} {\bibfnamefont
  {G.}~\bibnamefont {Petersson}},  \emph {et~al.},\ }\href@noop {} {\bibfield
  {journal} {\bibinfo  {journal} {Wallingford, CT}\ } (\bibinfo {year}
  {2009})}\BibitemShut {NoStop}%
\bibitem [{\citenamefont {Cirera}\ \emph {et~al.}(2018)\citenamefont {Cirera},
  \citenamefont {Via-Nadal},\ and\ \citenamefont
  {Ruiz}}]{cirera2018benchmarking}%
  \BibitemOpen
  \bibfield  {author} {\bibinfo {author} {\bibfnamefont {J.}~\bibnamefont
  {Cirera}}, \bibinfo {author} {\bibfnamefont {M.}~\bibnamefont {Via-Nadal}}, \
  and\ \bibinfo {author} {\bibfnamefont {E.}~\bibnamefont {Ruiz}},\ }\href@noop
  {} {\bibfield  {journal} {\bibinfo  {journal} {Inorganic chemistry}\ }\textbf
  {\bibinfo {volume} {57}},\ \bibinfo {pages} {14097} (\bibinfo {year}
  {2018})}\BibitemShut {NoStop}%
\bibitem [{\citenamefont {{Jmol development team}}()}]{jmol}%
  \BibitemOpen
  \bibfield  {author} {\bibinfo {author} {\bibnamefont {{Jmol development
  team}}},\ }\href {http://www.jmol.org/} {\enquote {\bibinfo {title} {Jmol: an
  open-source java viewer for chemical structures in 3d.}}\ }\BibitemShut
  {NoStop}%
\end{thebibliography}



%

\end{document}